# Deep learning-based image super-resolution of a novel end-expandable optical fiber probe for application in esophageal cancer diagnostics

Xiaohui Zhang,[a] Mimi Tan,[b] Mansour Nabil,[b] Richa Shukla,[b] Shaleen Vasavada,[b] Sharmila Anandasabapathy,[b,c] Mark A. Anastasio,[a*] Elena Petrova[b,c*]

[a]Department of Bioengineering, University of Illinois Urbana-Champaign, Urbana, IL, USA 61801
[b]Section of Gastroenterology and Hepatology, Department of Medicine, Baylor College of Medicine, TX, USA 77030
[c]Baylor Global Health, Baylor College of Medicine, TX, USA

**Abstract**

**Significance:** Endoscopic screening for esophageal cancer may enable early cancer diagnosis and treatment. While optical microendoscopic technology has shown promise in improving specificity, the limited field of view (<1 mm) significantly reduces the ability to survey large areas efficiently in esophageal cancer screening.

**Aim:** To improve the efficiency of endoscopic screening, we proposed a novel end-expandable endoscopic optical fiber probe for larger field of visualization and employed a deep learning-based image super-resolution (DL-SR) method to overcome the issue of limited sampling capability.

**Approach:** To demonstrate feasibility of the end-expandable optical fiber probe, DL-SR was applied on simulated low-resolution (LR) microendoscopic images to generate super-resolved (SR) ones. Varying the degradation model of image data acquisition, we identified the optimal parameters for optical fiber probe prototyping. The proposed screening method was validated with a human pathology reading study.

**Results:** For various degradation parameters considered, the DL-SR method demonstrated different levels of improvement of traditional measures of image quality. The endoscopists' interpretations of the SR images were comparable to those performed on the high-resolution ones.

**Conclusions:** This work suggests avenues for development of DL-SR-enabled end-expandable optical fiber probes to improve high-yield esophageal cancer screening.

*Elena Petrova, Email: elena.petrova.biocryst@gmail.com

## 1. Introduction

With an incidence of about 600,000 new cases and more than 500,000 deaths annually worldwide, esophageal cancer (EC) is one of the deadliest cancers worldwide. The two main histological subtypes of EC include: esophageal adenocarcinoma (EAC)[1] and esophageal squamous cell carcinoma (ESCC)[2]. Screening and early detection of ESCC are critically important to help reduce the incidence and mortality associated with EC. However, challenges posed by the extensive surface area of the esophagus, the limited field-of-view (FOV) offered by conventional microendoscopic screening probes (100-1000 µm) [3–5], and the time constraints of endoscopic procedures (6-8 minutes per patient) [1], necessitate the development of an innovative approach to enable imaging of larger fields and improve diagnostic yield.

In recent years, in-vivo optical microscopy techniques such as confocal laser endomicroscopy (CLE) and high-resolution microendoscopy (HRME), have been applied to visualize the nuclear morphology of the esophageal epithelium and assist in differentiating neoplasia from benign tissue [3,5–7]. In contrast to conventional tissue biopsies and histopathologic analysis, these optical techniques are non-invasive and able to provide real-time results. However, microendoscopy is currently limited by its small FOV (the average diameter of a probe is 0.3 mm) that hinders it from imaging larger tissue areas.

To circumvent this limitation, we propose a novel approach using an end-expandable optical probe to increase the FOV using sparse-data imaging [8,9]. By use of a sleeve mechanism that expands the end of the fiber bundle with unfused microfibers, a larger tissue FOV can be achieved for real-time diagnosis (Figure 1). However, one major challenge for successful application of such an optical probe is the development of image processing procedures that act on the acquired sparse image data and allow formation of a microendoscopic image with enhanced quality. Motivated by recent advances of image super-resolution using deep learning approaches [8,10–12], here we proposed a deep learning-based image super-resolution (DL-SR) method that estimates a HRME image from its acquired low-resolution (LR) counterpart. In this way, sparse images can be acquired by the end-expandable optical fiber probe to increase the FOV at cost of spatial resolution, which could be subsequently restored by use of a DL-SR method to enhance image quality (IQ) for diagnostic purposes.

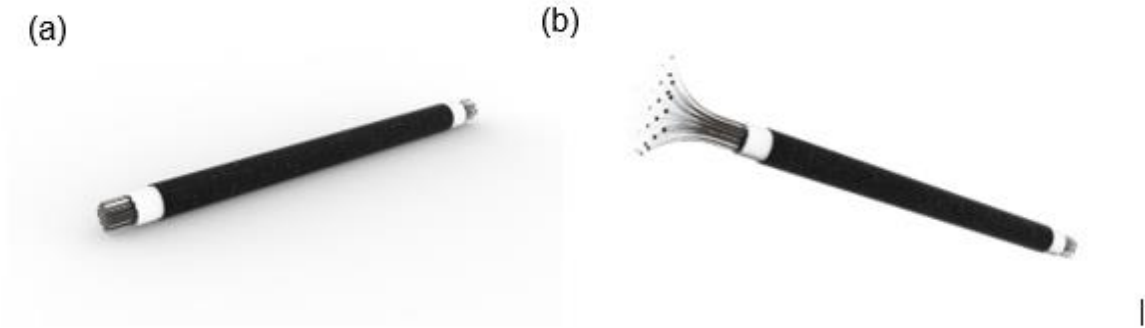

**Figure 1.** Conceptual 3D rendering of a settled optical fiber probe (a) and an end-expanded optical fiber probe (b), respectively. Here, the diameter of a single fiber strand can vary from 4 to 6 µm.

The goal of our study is to demonstrate the feasibility of the end-expandable optical fiber probe and the proposed sparse imaging methodology to achieve a wider FOV with image quality comparable to that achieved by a conventional probe of fused fiber. As a surrogate of clinical trials, computer simulation studies, also known as virtual imaging trials, provide us with an economical and convenient route to explore imaging system designs [13]. Here, estimates of LR microendoscopic images of esophageal mucosa that would be acquired by use of the novel optical fiber probe are computationally simulated. This will be accomplished by degrading HRME images by use of a degradation model that incorporates the physical factors of the optical fiber probe. Subsequently a DL-SR model was employed to generate super-resolved (SR) images from the LR ones. The impact of various degradation parameters of the probe on DL-SR performance was investigated to identify optimal parameters for prototyping. Additionally, a clinically-relevant detection task of esophageal neoplasia was conducted by endoscopists to study the impact of DL-SR on task performance. The results will provide valuable guidance for future prototyping and the advancement of this novel imaging technique.

## 2. Methods

To demonstrate the applicability of the proposed end-expandable optical fiber probe model, DL-SR models were trained in an end-to-end manner to learn a mapping between the LR images simulated with degradation models that incorporate different optical probe parameters and the HRME images. Traditional IQ metrics were computed on simulated LR and SR images to assess the perceptual image quality and identify the optimal parameters for the probe prototyping. In addition, a human pathology reading study was carried out to evaluate the utility of the images enhanced by the DL-SR model within a screening context.

*2.1 HRME image acquisition and pathology interpretation*

A HRME image dataset acquired by use of a low-cost, point-of-care HRME device that imaged the esophageal epithelium of patients undergoing endoscopy for ESCC screening [14] was used in the study. The HRME images were obtained sequentially from patients enrolled in a clinical trial comparing standard of care Lugol's chromoendoscopy (LCE) to LCE+HRME at 3 sites: First Hospital of Jilin University (Changchun, China), the Cancer Institute at The Chinese Academy of Medical Sciences (Beijing, China), and Baylor College of Medicine (Houston, Texas, USA) from December 2014 to November 2016, approved by the Institutional Review Board at Baylor College of Medicine [IRB# H-34973]. The details of image acquisition have been described in previous studies [14,15].

The HRME imaging system consisted of a compact epi-fluorescence fiber optic microscope which provides 1000x magnification views of epithelial tissue and subcellular features to distinguish cancerous from benign tissue after staining with a topical fluorescent dye, 0.01% proflavine hemisulfate. It has been evaluated in various anatomical sites, such as the cervix, anus, mouth, throat, and esophagus [5,15]. Due to its low cost and reusability, HRME can have a high impact in resource-limited global settings. The HRME principles and schematics have been described previously [14]. Typically, the HRME device conducts imaging through a fused optical fiber bundle with a cross-sectional diameter of 800 μm and 4.4 μm lateral spatial resolution (an individual fiber strand diameter as little as 4 um). For a conventional microscopic probe, the input and the output at the distal and proximal ends of the fibers demonstrate the same pattern. The fiber strands are positioned as close to each other as possible, and the size of the fiber strands usually defined image resolution for a registering camera of high-resolution. The light illumination measures are collected from a continuous part of tissue and registered on a camera correspondingly. Figure 2a demonstrates the optical fiber pattern generated by the fiber strands of the fused optical fiber bundle.

The density of the nuclei represented the metric for diagnosis and differentiation between neoplastic and non-neoplastic images. All image sites were biopsied, and biopsy histopathology results served as the gold standard. High-grade dysplasia and ESCC were classified as neoplastic; normal squamous epithelium, esophagitis, and low-grade dysplasia were classified as non-neoplastic.

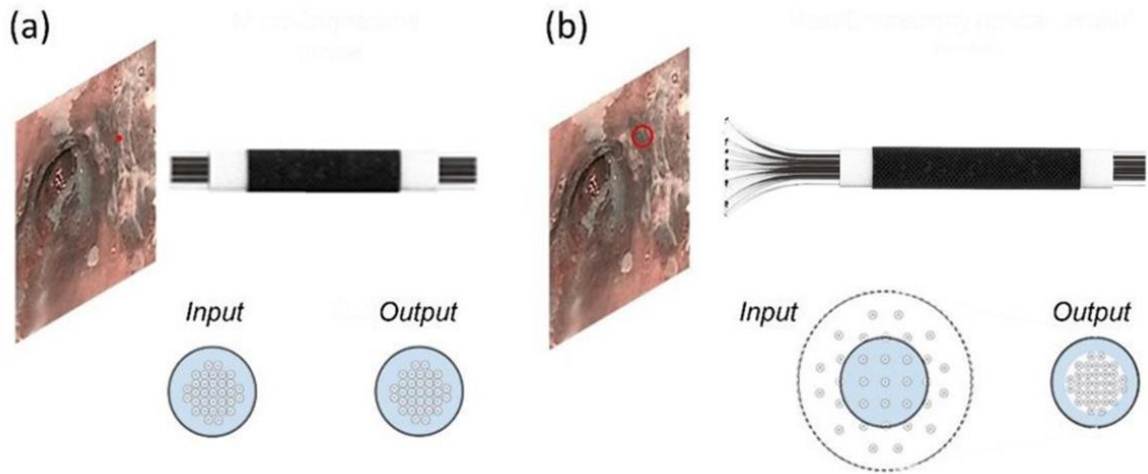

**Figure 2.** Schematics of image acquisition by means of a) a conventional fused endoscopic fiber probe; (b) an end-expandable, unfused endoscopic fiber probe.

All HRME images were standardized in size of 960 pixels × 1280 pixels. A Gaussian filter with standard deviation of 2 pixels was applied to remove the comb pattern introduced by the sparse data of optical fiber bundles. A contrast-limited adaptive histogram equalization (CLAHE) with a clip limit of 0.005 was employed to enhance the image contrast.

*2.2 Quality control and image selection*

In addition to image quality control conducted as described in the previous study[15], bioengineers further categorize the acquired HRME images into three types of perceptual image qualities: good, intermediate, and poor. Images with good quality should demonstrate a FOV where nuclei can be clearly visualized. Images with intermediate quality showed mild motion blur or defocus aberration that could affects approximately a quarter of the imaged area. Images with poor quality were classified by factors such as severe motion blur, obstructed vision or image corruption for half of the image area or more. Only images of good and intermediate perceptual quality were selected and used in this study.

*2.3 Simulated image data for end-expandable optical fiber probe*

A virtual imaging trial was performed to simulate LR images that would be acquired using the proposed novel optical fiber probe by use of the acquired HRME images. As shown in Figure 2b, one can find a schematic of the concept to an end-expandable optical fiber probe. Unlike conventional image acquisition through an optical fiber bundle (Figure 2a), the end-expandable optical fiber bundle can form a "brush" at the input end (Figure 2b). The input elements of such lightguide are fiber strands which are placed at a distance from each other and all together they represent a sparse data set. Thus, the optical fiber "brush" collects fluorescent emission from a surface partially and discontinuously. To create LR images, we processed the simulated sparse image data acquired from the end-expandable endoscopic optical fiber probe (Figure 2b) by filling in non-populated pixels of the sparse images (Figure 3b). Diagrams that depict the conventional and sparse image data acquisition and processing are shown in Figure 3.

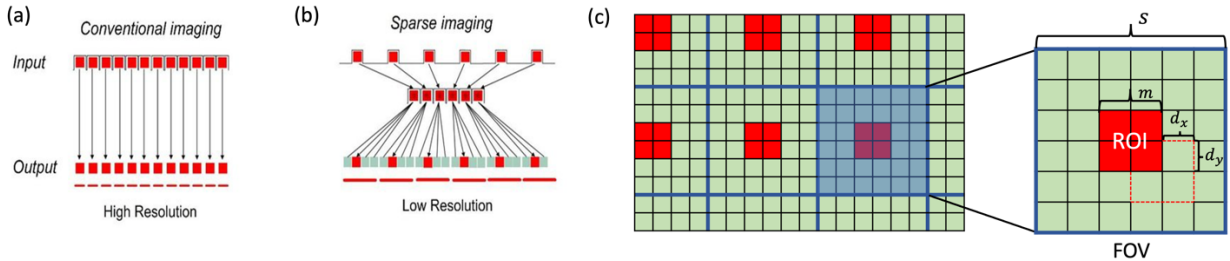

**Figure 3.** (a) 1D-schematic of the conventional image data acquisition leading to the HR images; (b) the sparse image data acquisition and reconstitution from the compressed data set into the LR images. Red elements represent a fiber strand collecting light illuminated by the tissue surface and green elements are a set of pixels displayed as average value of light intensity in the adjacent fiber. (c) 2D-illustration of the simulated sparse image data; $s$ and $m$ are the length of side of FOV and ROI, respectively; $d_x$ and $d_y$ are the offsets in $x$ and $y$ directions, respectively.

To emulate the LR images that would be acquired from the novel end-expandable optical fiber probe, a degradation model that incorporates the fiber knock-out downsampling was applied to the HRME images. Assuming a single fiber strand with a diameter of 4 μm, we considered a pixel size of 2 μm in this study. As shown in Figure 3, given a fiber strand represented by a range-of-interest (ROI) block of $m$ pixels × $m$ pixels that resides in the center of a FOV block of $s$ pixels × $s$ pixels, a random deformation offset with maximum of $d_x$ pixels and $d_y$ pixels in horizontal and vertical direction was considered, respectively. This was designed to depict the off-center deviation when defining the ROI block. The degradation model calculated the mean value over pixels within the ROI block and filled the pixels in FOV block with the derived value. In this study, three different parameters of the degradation model were investigated to evaluate the DL-SR performance: the fiber diameter $m$, the inter-fiber distance $s$, and the deformation offset $d$. The schematic of the proposed degradation model is shown in Figure 4.

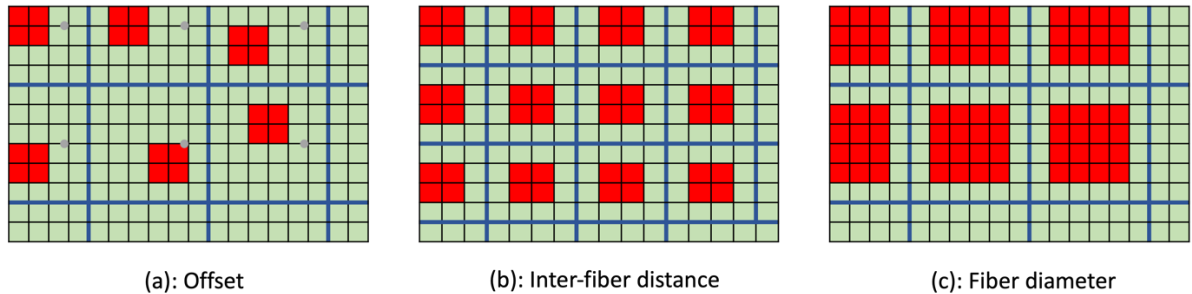

(a): Offset     (b): Inter-fiber distance     (c): Fiber diameter

**Figure 4.** Illustration of degradation models that incorporated various optical fiber probe parameters including (a) offset, (b) inter-fiber distance and (c) fiber diameter.

Figure 5 shows examples of an HRME image, an intermediate sparse image and the LR image simulated by use of the degradation model, respectively. Here, the fiber diameter $m = 6$ µm, the inter-fiber distance $s = 12$ µm, and the deformation offset $d = 1$.

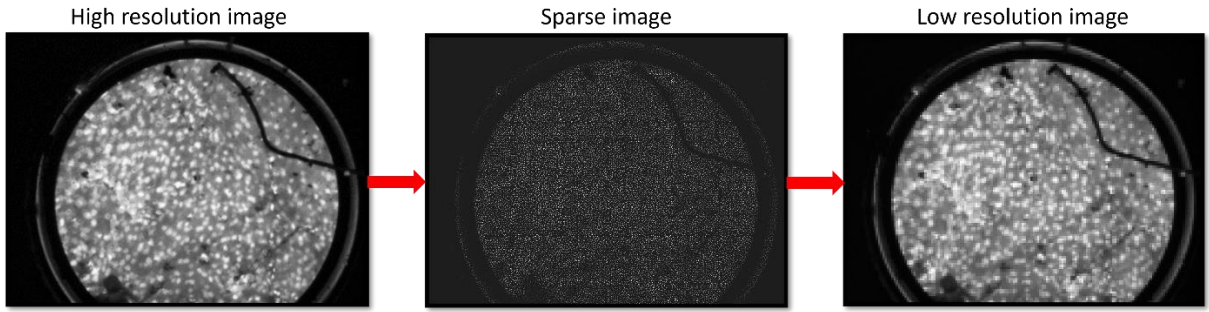

High resolution image     Sparse image     Low resolution image

**Figure 5.** Example of an original HRME image, the same imaging area as sparse data captured by the proposed end-expandable optical fiber probe, and the LR image restored from the sparse data.

*2.4 Deep learning-based image super-resolution*

Given an LR image $\mathbf{I}_{LR}$ virtually acquired with the end-expandable optical probe, image super-resolution methods seek to produce a super-resolved image $\mathbf{I}_{SR}$ as an estimate of a HRME image that would be obtained from a conventional fused endoscopic fiber probe. However, this is a challenging ill-posed inverse problem. In recent years, deep learning-based super-resolution methods have been widely applied in various applications [10–12,16–20]. Here, the well-studied super-resolution convolutional neural network (SRCNN) [10] was employed to investigate the feasibility of DL-SR methods to improve quality of simulated LR images from the novel optical probe. Such analysis can be readily repeated with other more recent DL-SR approaches[11,21]. The SRCNN seeks to establish a mapping from the space of simulated LR images to the space of HR images:

$$\mathbf{I}_{SR} = \mathcal{F}(\mathbf{\Theta}, \mathbf{I}_{LR}),$$

where $\mathcal{F}$ is the network parameterized by $\mathbf{\Theta}$. The SRCNN was trained by minimizing the mean squared error (MSE) between generated SR images and original HRME images. The architecture of a SRCNN is shown in Figure 6, consisting of three feedforward convolutional layers interspersed with leaky rectified linear unit (LReLU) nonlinearities. The filter sizes of the three convolutional layers were 9×9, 1×1, and 5×5 and the corresponding number of filters were 64, 32, and 1, respectively.

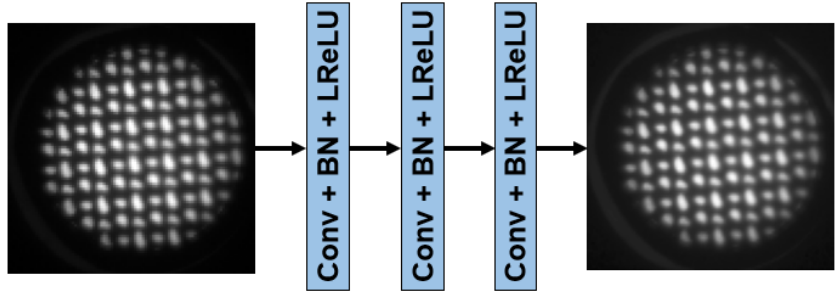

**Figure. 6** Schematic of the SRCNN architecture used in the study.

The training and validation data for the SRCNN consisted of 206 and 50 paired HRME and corresponding simulated LR images, respectively. During training and validation, each image was randomly cropped into 10 patches with a size of 512 × 512. In the testing stage, the full-size simulated LR images were used. The SRCNN was trained with Adam optimizer [22] with a learning rate of 0.0001. The network was trained for 300 epochs with a batch size of 8, and the model with best validation performance was used for downstream task evaluation. For various degradation parameters mentioned in Section 2.3, the SRCNN was retrained and evaluated. SRCNN were implemented with TensorFlow 2.0 and trained on NVIDIA GPUs.

*2.5 Image quality assessment and statistical analysis*

To assess the DL-SR performance with consideration of various degradation parameters used to simulate the LR images, traditional IQ metrics such as peak-signal-to-noise ratio (PSNR) and structural similarity index metric (SSIM) were computed on the simulated LR and SR images. In addition, a binary detection task to determine whether esophageal neoplasia is present or not was performed by endoscopists on both SR images obtained from SRCNN and the original HRME ones. The accuracy and confidence level were evaluated to assess the task-based performance of the employed DL-SR method.

A total of 4 endoscopists (3 experts, 1 novice) underwent standardized training in HRME image interpretation. Expert endoscopists were defined as having previously performed >50 HRME cases, whereas novices were new to the technology. All endoscopists viewed a set of training slides that demonstrated the features of neoplastic and non-neoplastic classification of HRME images, including nuclear size, crowding, and pleomorphism. All endoscopists were asked to interpret a series of original HRME images and generated SR images as neoplastic (high-grade dysplasia, ESCC) or non-neoplastic (normal squamous, esophagitis, low-grade dysplasia) along with their confidence level in their interpretation (high or low).

Assuming 80% power, $\alpha = 0.05$, and 2-sided test to establish equivalence between HR and SR images with equivalence limit of 0.15, a sample size of 120 images was calculated using sample-based variance estimates without continuity correction for binary data. The sensitivity, specificity, and accuracy of HRME and SR image interpretation of individual endoscopists was compared with histopathology as the gold standard. Unpaired t-test was computed to assess for differences in the sensitivity, specificity, and accuracy in endoscopists' interpretation of HR and generated SR images.

## 3. Results

*3.1 Performance evaluation of DL-SR to super-resolve simulated LR images of end-expandable optical fiber probe*

The DL-SR model was first trained and tested on HRME images and corresponding LR images simulated by use of a degradation model with fiber diameter m=4 μm, inter-fiber distance s=8 μm and no offset imposed. Figure 7 shows two examples of the test images where perceptual image quality of the simulated LR images appeared to be improved by DL-SR. Figure 8 shows a zoomed-in, cutoff image patch of the original HR, simulated LR, and generated SR images and the corresponding cross-sectional profiles of optical intensity. Note that, as expected, the line profiles of the SR image aligned closer to the HR one, compared to the line profiles of the simulated LR image. Visual inspection confirms that details of subcellular features could be restored by use of SRCNN to obtain improved perceptual image quality.

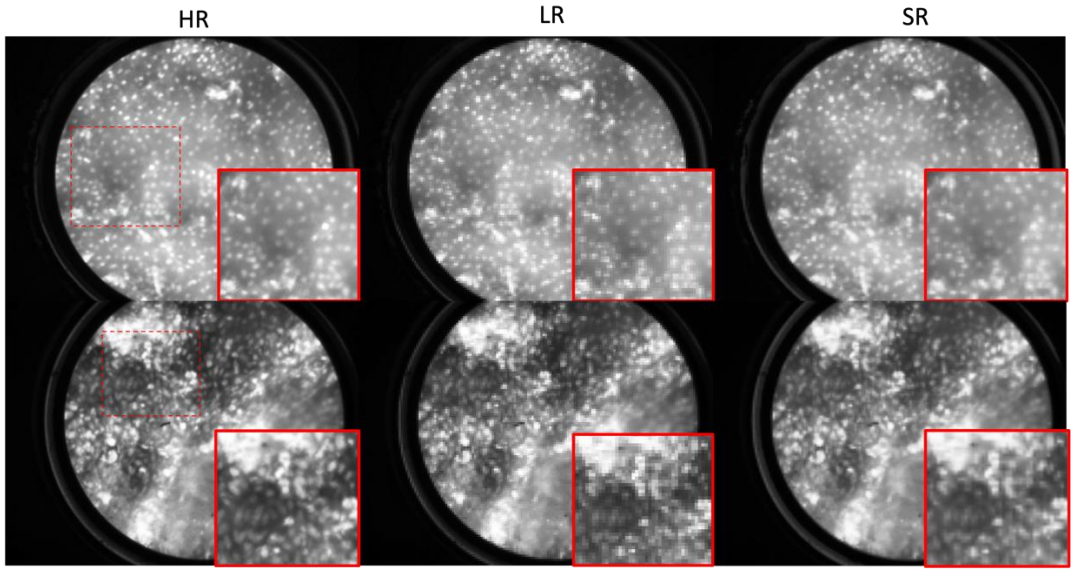

**Figure 7.** Examples of original HRME image, simulated LR image and corresponding SR image (top row) generated by the DL-SR method with magnification (bottom row).

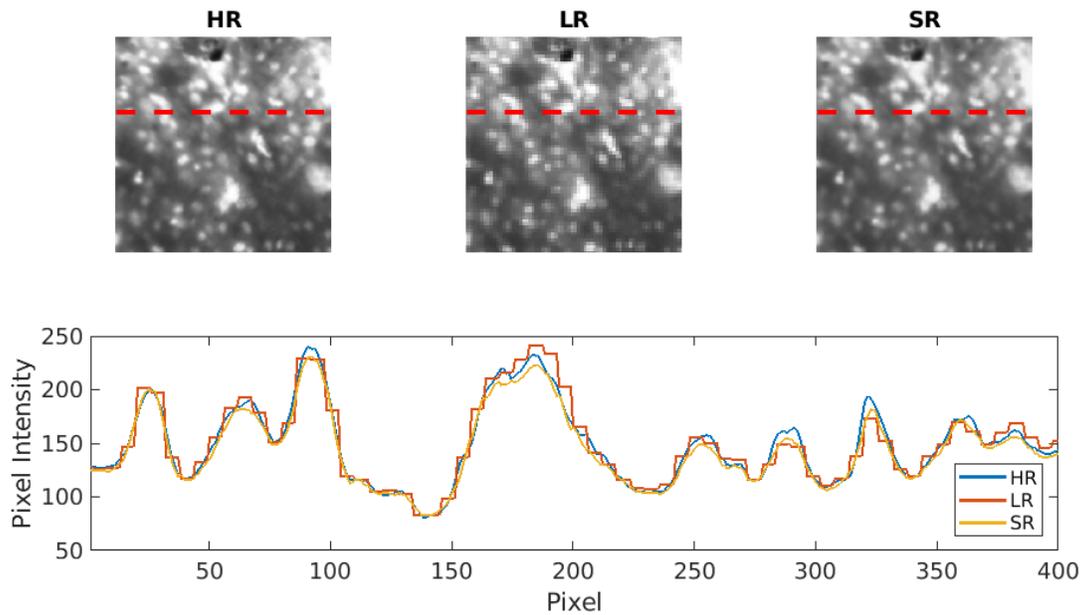

**Figure 8.** Cross-sectional line profile on selected images allows comparison of profiles of optical intensity associated with stained nuclei on original HR images, simulated LR images, and reconstructed SR images.

*3.2 Impact of offset, inter-fiber distances and fiber diameter on traditional image quality metrics*

In this study, the impact of different parameters considered in the degradation model on traditional IQ metrics was evaluated. The offset $d$ was varied from 0 μm to 10 μm, and the inter-fiber distance $s$ was changed ranging between 4 μm to 24 μm. A fiber strand diameter $m$ ranging from 4 μm to 12 μm was applied. All SRCNN models were trained on paired of HRME and LR images simulated with various degradation parameter values. Traditional measures of image quality were assessed by computing PSNR and SSIM values on a test set consisting of 300 images, and these quantities are plotted on Figure 9. In most cases, the SR images generated by the SRCNN demonstrated improvements across various offsets, inter-fiber distances and fiber diameters compared with their LR counterparts in terms of the traditional IQ metrics. Moreover, various degradation parameters showed different levels of impacts on the traditional IQ metrics of simulated LR and SR images. When increasing the offset of fibers, traditional IQ metrics of both LR and SR images decreased. Similar phenomena were observed when the inter-fiber distance and fiber diameters were increased. This is due to the more severe degradation model incorporated and thus DL-SR performs worse on images largely missing information when the network capacity or the number of training images were limited.

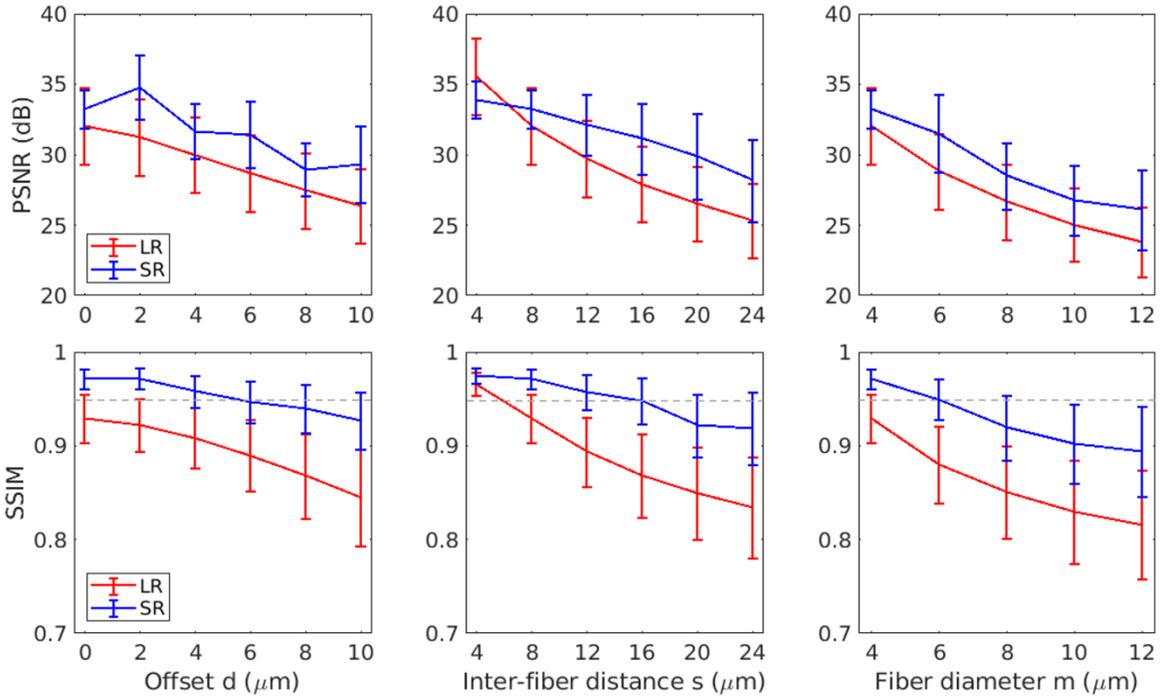

**Fig. 9.** Traditional IQ metrics show degradation of LR and limitation of improvement of SR images with reference to HR images: PSNR and SSIM values of LR (red) and SR (blue). The gray dashed line denotes a SSIM value of 0.95.

*3.3 Endoscopists' HRME image interpretation of DL-SR images*

Among the 120 paired HRME images and SR images included in the study, 42 images (35%) were non-neoplastic and 78 images (65%) were neoplastic. The diagnostic performance of endoscopists on HR and SR images is demonstrated in Figure 10 and 11. Overall, the task-based image quality of SR images was comparable to that of HR images when endoscopists interpreted them in a post-hoc setting. On the original HR images, the endoscopists achieved an overall accuracy of 70.2% (standard deviation [SD], 5.4%), a sensitivity of 73.7% (SD, 14.1%), and a specificity of 63.7% (SD, 12.8%). On the SR images, the endoscopists did not significantly change their accuracy (p=0.37), sensitivity (p=0.70), or specificity (p=0.07).

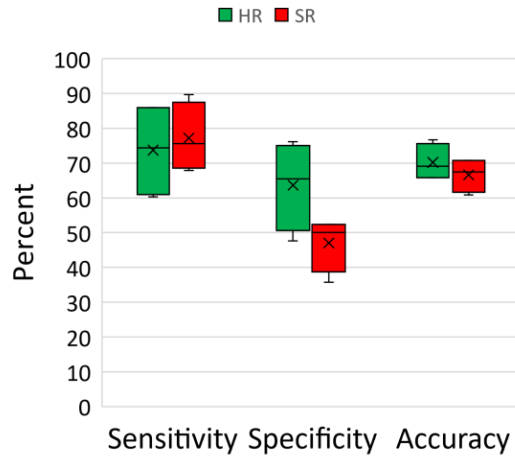

**Figure 10.** Diagnostic performance of endoscopists' reading on HR and SR microendoscopy images shown in box and whisker plot. The cross and horizontal line in the box represents mean and median values of the endoscopists' response. The bottom and the top end from the whiskers are the minimum and maximum values, respectively.

Overall, endoscopists had high confidence in their image interpretation in a mean 51.1% (SD, 7.7%) of HR images and 48.5% of SR images (SD, 9.0%, p=0.68). Among high confidence images, endoscopists had no significant difference in their accuracy (76.0% vs. 74.8%, p=0.73), sensitivity (84.6% vs. 88.3%, p=0.45), or specificity (45.9% vs. 30.2%, p=0.19) on SR images compared to HR counterparts. Similarly, for low-confidence images, endoscopists did not have a significant difference in their accuracy (64.2% vs. 60.1%, p=0.39), sensitivity (58.1% to 63.7%, p=0.68), or specificity (73.3% vs. 55.8%, p=0.14) on SR images compared to HR images.

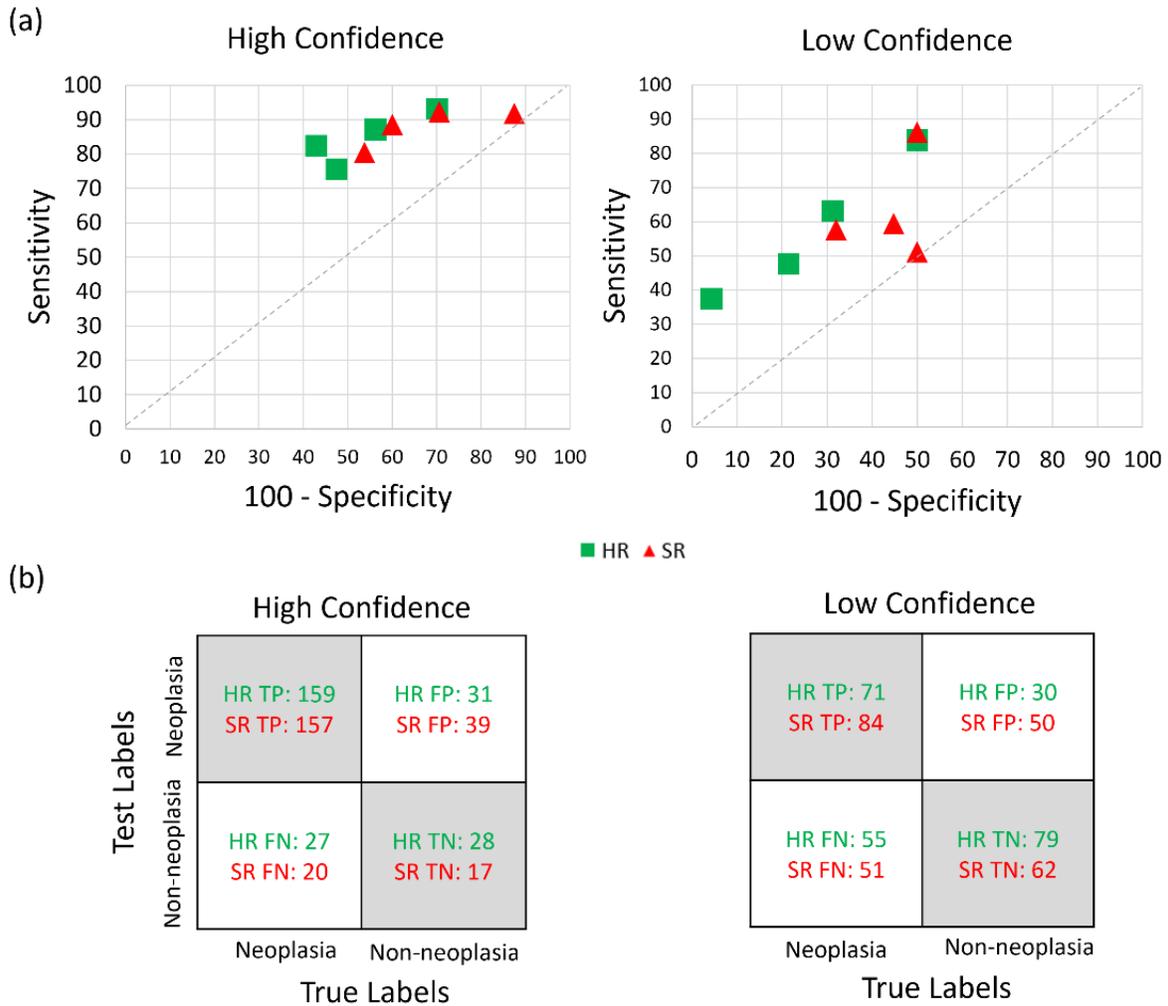

**Figure. 11** (a) Sensitivity and specificity plot of individual endoscopist diagnosis on HR and SR microendoscopic images with high and low confidence, respectively (b) Confusion matrix of diagnostic performance of readers on HR and SR images with low and high confidence, respectively. TP, true positive; FP, false positive; FN, false negative; TN, true negative.

## 4. Discussion

In this study, an endoscopic end-expandable optical fiber probe was proposed and evaluated for its ability to improve the field of imaging in esophageal cancer screening. The novel concept of an end-expandable optical fiber probe was simulated by use of degradation model incorporating various probe parameters and was tested using HRME images of esophageal squamous tissue. We performed a virtual imaging trial to simulate LR microendoscopic image dataset from clinical HRME images and demonstrated the effectiveness of a deep learning-based super-resolution algorithm to improve the perceptual image quality measures of LR images. Furthermore, the generated SR images were comparable to the original HRME images when interpreted by endoscopists in a diagnostics task.

The capability of the employed DL-SR algorithm to super-resolve a LR image of end-expandable optical fiber probe was proven for the first time here and should be highlighted because of its potential for further improvement and application. As expected, the quality of the SR images deteriorated when severe degradation models were considered. This indicated the significance of choosing tolerable parameters for future physical prototypes of the end-expandable optical fiber probe. We found that the diameter of a single fiber strand should be no more than 6 µm. The inter-fiber distance can be increased to up to 14-16 µm which should be taken into account in the design and development of an optical fiber probe sleeve and a mechanism for controlling the inter-fiber distance. The fiber flexibility can permit offset of up to 4-6 µm. As a result, it will allow for increasing the diameter of an endoscopic sensor up to 4-5 mm instantly (the FOV is 10-20 folds bigger than HRME) that can provide a dramatic improvement in esophageal screening efficiency.

Endoscopists had comparable diagnostic accuracy when interpreting original HRME images and DL-SR generated SR ones. While specificity could be improved, sensitivity was higher for SR images than for the original HR images. Our metrics of success focused on achieving comparable performance of human reads on SR images compared to that on HR images, which was achieved. Moreover, endoscopists had comparable level of confidence when interpreting SR images compared to the original HR images, which is a quantitative task-based assessment of the SR image quality. DL-SR methods that effectively estimate SR images, resembling the original

HR images, is a robust strategy to increase the field of visualization during esophageal cancer screening while maintaining accurate esophageal neoplasia detection by clinicians. More advanced deep learning-based image super-resolution methods that employ generative adversarial networks [11,19] and diffusion models [21] should also be investigated for such application in future studies.

This study evaluated the feasibility, imaging capability, and clinical performance of a hypothetical end-expandable optical fiber probe based on simulation studies using existing HRME images. Future *in vivo* studies will need to develop and test the fiber probe and its effectiveness in esophageal cancer imaging. Using a larger dataset of the clinical microendoscopic images will potentially further advance the DL-SR performance. Another limitation of this study includes the rectangular orientation of the degradation model. A radial model of the optical fiber probe expansion will be applied to future studies.

## 5. Conclusion

We proposed an end-expandable optical fiber probe that would increase the field of visualization by up to 20-fold compared to a traditional fused HRME probes. We further validated the DL-SR generated images produced from the end-expandable optical fiber probe and found endoscopists' interpretation of the SR images to be comparable to that of conventional HRME images. The proposed novel end-expandable optical fiber probe has the *potential* to enable high-yield endoscopic microscopy screening and even facilitate a screen-and-treat protocol for early esophageal cancer treatment. Moreover, the proposed sparse image methodology will provide valuable guidance for future prototyping and the advancement of optical biopsy techniques over larger surface areas.

*Disclosure*

The authors declare no potential conflicts of interest.

*Acknowledgments*

The authors acknowledge the contributions of Drs. Rebecca Richards-Kortum and Richard Schwarz from Rice University (Houston, TX) who provided the HRME instruments and helped to collect the clinical trial HRME images utilized here and who provided advice on the critical features of HRME images for data evaluation and image processing. Additionally, to the endoscopists of Baylor College of Medicine, a part of the HRME images were collected by Drs. Guiqi Wang, Fan Zhang, Hong Xu (Beijing and Changchun, China) whom we are grateful as well.

This work was supported in part by the National Institutes for Health, National Cancer Institute, Award Nos. R01 CA181275. Dr. Tan is supported by the National Institute of Diabetes and Digestive and Kidney Diseases of the National Institutes of Health under Award Number K23DK129776. Dr. Mark A. Anastasio was supported in part by NIH Award EB031772 (subproject 6366). Preliminary results of this work were presented at SPIE Photonics West 2021 and published as an SPIE Proceedings paper.

*References*

1. N. M. Mansour, S. S. Groth, and S. Anandasabapathy, "Esophageal Adenocarcinoma: Screening, Surveillance, and Management," Annual Review of Medicine **68**(1), 213–227 (2017) [doi:10.1146/annurev-med-050715-104218].
2. J. Yang et al., "Understanding Esophageal Cancer: The Challenges and Opportunities for the Next Decade," Frontiers in Oncology **10** (2020).

3. B. S, "Progress and Challenges of Global High-Resolution Endoscopy," clinmed journals [doi:10.23937/2643-4466/1710024].

4. P. Sharma et al., "Real-time increased detection of neoplastic tissue in Barrett's esophagus with probe-based confocal laser endomicroscopy: final results of an international multicenter, prospective, randomized, controlled trial," Gastrointest Endosc **74**(3), 465–472 (2011) [doi:10.1016/j.gie.2011.04.004].

5. Y. Tang, S. Anandasabapathy, and R. Richards-Kortum, "Advances in optical gastrointestinal endoscopy: a technical review," Mol Oncol **15**(10), 2580–2599 (2021) [doi:10.1002/1878-0261.12792].

6. S. Bhushan et al., "Photoacoustic Imaging in Gastroenterology: Advances and Needs," in Photoacoustic Imaging - Principles, Advances and Applications, IntechOpen (2019) [doi:10.5772/intechopen.86051].

7. A. L. Polglase et al., "A fluorescence confocal endomicroscope for in vivo microscopy of the upper- and the lower-GI tract," Gastrointest Endosc **62**(5), 686–695 (2005) [doi:10.1016/j.gie.2005.05.021].

8. S. Anandasabapathy et al., "An optical, endoscopic brush for high-yield diagnostics in esophageal cancer," in Endoscopic Microscopy XVI **11620**, p. 116200B, SPIE (2021) [doi:10.1117/12.2583301].

9. E. Petrova, S. ANANDASABAPATHY, and R. Richards-Kortum, "End expandable optical fiber bundle," WO2022051741A2 (2022).

10. C. Dong et al., "Image Super-Resolution Using Deep Convolutional Networks," IEEE Transactions on Pattern Analysis and Machine Intelligence **38**(2), 295–307 (2016) [doi:10.1109/TPAMI.2015.2439281].

11. C. Ledig et al., "Photo-Realistic Single Image Super-Resolution Using a Generative Adversarial Network," presented at Proceedings of the IEEE Conference on Computer Vision and Pattern Recognition, 2017, 4681–4690.

12. W.-S. Lai et al., "Fast and Accurate Image Super-Resolution with Deep Laplacian Pyramid Networks," arXiv:1710.01992 [cs] (2018).

13. S. Park et al., "Stochastic three-dimensional numerical phantoms to enable computational studies in quantitative optoacoustic computed tomography of breast cancer," JBO **28**(6), 066002, SPIE (2023) [doi:10.1117/1.JBO.28.6.066002].

14. M.-A. Protano et al., "Low-Cost High-Resolution Microendoscopy for the Detection of Esophageal Squamous Cell Neoplasia: An International Trial," Gastroenterology **149**(2), 321–329, Elsevier (2015) [doi:10.1053/j.gastro.2015.04.055].

15. M. C. Tan et al., "Automated software-assisted diagnosis of esophageal squamous cell neoplasia using high-resolution microendoscopy," Gastrointestinal Endoscopy **93**(4), 831-838.e2 (2021) [doi:10.1016/j.gie.2020.07.007].

16. Z. Wang, J. Chen, and S. C. Hoi, "Deep learning for image super-resolution: A survey," IEEE Transactions on Pattern Analysis and Machine Intelligence, IEEE (2020).

17. W. Yang et al., "Deep Learning for Single Image Super-Resolution: A Brief Review," IEEE Transactions on Multimedia **21**(12), 3106–3121 (2019) [doi:10.1109/TMM.2019.2919431].

18. X. Zhang et al., "Impact of deep learning-based image super-resolution on binary signal detection," JMI **8**(6), 065501, SPIE (2021) [doi:10.1117/1.JMI.8.6.065501].

19. S. Sengupta et al., "DeSupGAN: Multi-scale Feature Averaging Generative Adversarial Network for Simultaneous De-blurring and Super-Resolution of Retinal Fundus Images," in Ophthalmic Medical Image Analysis, H. Fu et al., Eds., pp. 32–41, Springer International Publishing, Cham (2020) [doi:10.1007/978-3-030-63419-3_4].

20. V. A. Kelkar et al., "Task-based evaluation of deep image super-resolution in medical imaging," in Medical Imaging 2021: Image Perception, Observer Performance, and Technology Assessment **11599**, pp. 207–213, SPIE (2021) [doi:10.1117/12.2582011].

21. C. Saharia et al., "Image Super-Resolution via Iterative Refinement," arXiv:2104.07636, arXiv (2021) [doi:10.48550/arXiv.2104.07636].

22. D. P. Kingma and J. Ba, "Adam: A Method for Stochastic Optimization," arXiv:1412.6980 [cs] (2017).